\documentclass[10pt,conference]{IEEEtran}  
\IEEEoverridecommandlockouts

\usepackage[ruled,linesnumbered]{algorithm2e} 
\usepackage{amsmath,amssymb,amsfonts} 
\usepackage{tabularx,booktabs,multirow} 
\usepackage{graphicx} 
\usepackage{cite} 
\usepackage{balance}
\usepackage[numbers,sort&compress]{natbib}
\usepackage{textcomp}
\usepackage{xcolor}
\usepackage{placeins} 
\usepackage{pdflscape} 

\def\BibTeX{{\rm B\kern-.05em{\sc i\kern-.025em b}\kern-.08em
    T\kern-.1667em\lower.7ex\hbox{E}\kern-.125emX}}  

\begin{document}

\title{Investigating and Mitigating Barren Plateaus in Variational Quantum Circuits: A Survey}

\author{\IEEEauthorblockN{
Jack Cunningham\textsuperscript{1}  \ \
Jun Zhuang*\textsuperscript{1}  \ \
}
\IEEEauthorblockA{
\textsuperscript{1}Boise State University, ID, USA.\\
{\scriptsize *Corresponding authors: junzhuang@boisestate.edu}
}
}



\maketitle  

\begin{abstract}
In recent years, variational quantum circuits (VQCs) have been widely explored to advance quantum circuits against classic models on various domains, such as quantum chemistry and quantum machine learning. Similar to classic machine-learning models, VQCs can be trained through various optimization approaches, such as gradient-based or gradient-free methods. However, when employing gradient-based methods, the gradient variance of VQCs may dramatically vanish as the number of qubits or layers increases. This issue, a.k.a. Barren Plateaus (BPs), seriously hinders the scaling of VQCs on large datasets. To mitigate the barren plateaus, extensive efforts have been devoted to tackling this issue through diverse strategies. In this survey, we conduct a systematic literature review of recent works from both investigation and mitigation perspectives. Furthermore, we propose a new taxonomy to categorize most existing mitigation strategies into five groups and introduce them in detail. Also, we compare the concurrent survey papers about BPs. Finally, we provide insightful discussion on future directions for BPs.
\end{abstract}



\section{Introduction}
\label{sec:intro}

In the era of noisy intermediate-scale quantum (NISQ), quantum computing has achieved significant advancement in various domains~\cite{preskill2018quantum}, such as quantum chemistry~\cite{bauer2020quantum}, quantum machine learning~\cite{liang2022variational, thanasilp2023subtleties, zhang2024generative, vijendran2024expressive}, and quantum architecture~\cite{chiribella2008quantum, liang2023unleashing}. Despite these advancements, the superiority of quantum computing over classical computing in the foreseeable future remains uncertain. One uncertainty arises during the optimization of Variational Quantum Circuits (VQCs), which have emerged as pivotal models for advancing quantum computing. Similar to classical models, VQCs can be optimized by gradient-based approaches~\cite{cerezo2021cost, uvarov2021barren}. During optimization, however, the training of VQCs may be initially trapped in a flat landscape as the model size increases. McClean et al.~\cite{McClean2018landscapes} first categorize this issue as barren plateaus (BPs) in VQCs and ascertain that the variance of the gradient will exponentially decrease as the model size increases when the VQCs match the 2-design Haar distribution. Under this occurrence, most gradient-based techniques would fail. Thus, it is essential to mitigate the barren plateau issues in VQCs.

\begin{figure}[t]
    \centering
    \includegraphics[scale=0.6]{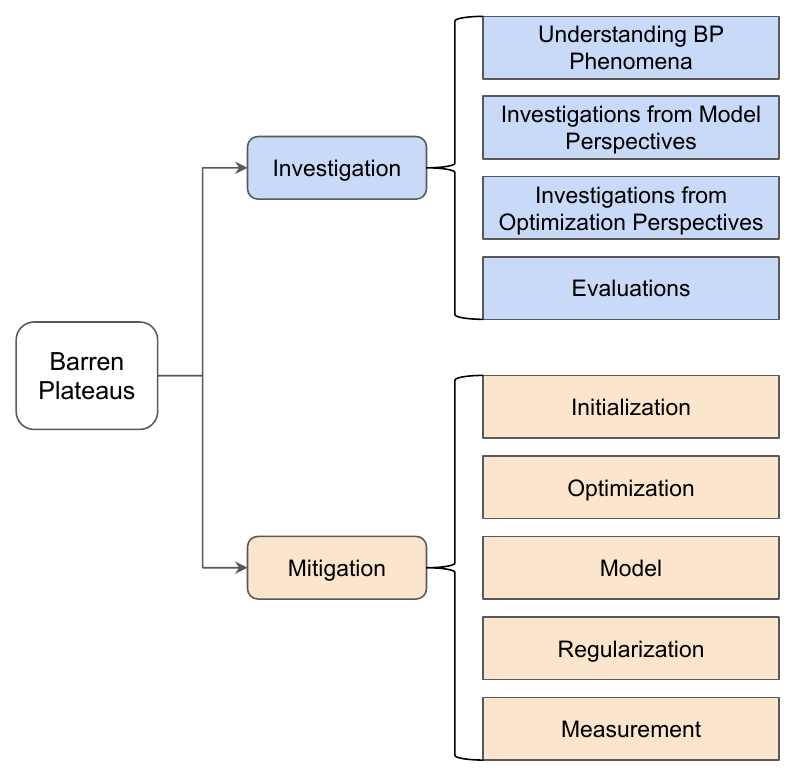}
    \caption{Our proposed taxonomy. We categorize the most existing works about barren plateaus into two aspects: investigation and mitigation.}
    \label{fig:taxonomy}
\end{figure}

Extensive efforts have been devoted to tackling the challenges of barren plateaus through various approaches~\cite{qi2023barren}. To categorize these studies, we conduct comprehensive literature reviews on recent works about BPs from two perspectives. On the one hand, some studies aim to investigate the barren plateau phenomenon. In this group, several works conduct experiments to understand and evaluate the barren plateaus. On the other hand, extensive strategies are proposed to mitigate the barren plateaus. To better categorize these works, we propose a new taxonomy presented in Figure~\ref{fig:taxonomy}. Overall, our contributions to this survey can be summarized as follows:
\begin{itemize}
  \item We conduct comprehensive literature reviews on recent works about barren plateaus (BPs)\footnote{Our repo: https://github.com/junzhuang-code/Barren\_Plateaus\_Survey/}.
  \item We propose a new taxonomy to help researchers better understand the background and existing works about BPs from two aspects: investigation and mitigation.
  \item We provide insightful discussions in this survey, including a summary of most existing mitigation methods, a comparison of concurrent survey papers, a contrast between gradient issues and barren plateaus, and a discussion of future directions.
\end{itemize}

In the rest part, we first introduce the preliminary background of BPs in Section~\ref{sec:background}. Besides, we present the investigation of BPs in Section~\ref{sec:investigations} and mitigation strategies in Section~\ref{sec:mitigations}. Last, we provide insightful discussion in Section~\ref{sec:discussion}.

\section{Preliminary of Barren Plateaus}
\label{sec:background}

The term, ``Barren Plateaus (BPs)", typically denotes an issue where gradient-based approaches may initially struggle to optimize Variational Quantum Circuits (VQCs). Generally, VQCs contain a finite sequence of unitary gates $U(\theta)$ parameterized by $\theta$. The unitary gates $U(\theta)$ can be formulated as:
\begin{equation}
    U(\theta) = U(\theta_1, ..., \theta_L) = \prod_{l=1}^{L} U_l(\theta_l),
\label{eqn:vqc}
\end{equation}
where $U_l(\theta_l) = e^{-i\theta_lV_l}$, $V_l$ is a Hermitian operator, and $L$ denotes the number of layers in VQCs.

Similar to classical models, VQCs' parameters can be optimized by gradient-based methods. The cost function for optimization can be denoted as $C(\theta)$. Thus, $C(\theta)$ can be formulated as the expectation over Hermitian operator $H$:
\begin{equation}
    C(\theta) = \langle0| U(\theta)^{\dagger} H U(\theta) |0\rangle.
\label{eqn:loss_fn}
\end{equation}

Based on the above loss function $C(\theta)$, we can further denote its gradient as $\frac{\partial C(\theta)}{\partial \theta_l}$, a.k.a. $\partial C$. Note that, in this work, we use the term ``optimization" to describe a process that searches optimal parameters for VQCs via numerical methods.

In 2018, McClean et al.~\cite{McClean2018landscapes} first investigated barren plateaus and revealed that under the assumption of the two-design Haar distribution, the variance of the gradient $Var[\partial C]$ in VQCs would exponentially decrease to zero during optimization as the model size, such as the number of qubits or layers, increases.
This phenomenon is rooted in the assumption of the Haar distribution, under which $U(\theta)$ is considered sufficiently random in Hilbert space, making the gradient landscape increasingly flat as the size of VQCs grows.
This randomness, a.k.a. Haar randomness, is characterized by the left- and right invariance of the Haar measure $\mu(U)$ on the unitary group $U(N)$, which ensures that the distribution of any function $f(U)$ remains unchanged under transformations by $V$ as follows.
\begin{equation}
 \int_{U(N)} d\mu(U) f(U) = \int d\mu(U) f(VU) = \int d\mu(U) f(UV).
\label{eqn:harr}
\end{equation}

The invariance presented in Equation~\ref{eqn:harr} plays a crucial role in ensuring the uniformity of random unitaries. To achieve such invariance, however, VQCs require large computational resources, which is impractical in both simulation and real-world scenarios. To mitigate this drawback, researchers often leverage unitary $t$-designs~\cite{renes2004symmetric} for restricted classes of $f(U)$, which significantly reduce computational demands while approximating the properties of the Haar measure.
Formally, suppose that we sample unitary $V_i$ with probability $p_i$, then a unitary ensemble $\{p_i, V_i\}$ is a $t$-design if it satisfies the following condition:
\begin{equation}
\sum_i p_i V_i^{\otimes t} \rho(V_i^\dagger)^{\otimes t} = \int d\mu(U) U^{\otimes t} \rho(U^\dagger)^{\otimes t},
\label{eqn:tdesign}
\end{equation}
where $\rho$ denotes the density matrix, and $\otimes$ denotes the tensor product.

Equation~\ref{eqn:tdesign} ensures that a $t$-design can replicate the statistical properties of Haar randomness for functions $f(U)$ of degree $t$ or lower. In practice, this means that if $f(U)$ is a polynomial with a maximum degree of $t$ in unitary $U$, averaging $f(U)$ over the $t$-design $\{p_i, V_i\}$ approximates the result of averaging it over the unitary group w.r.t. the Haar measure~\cite{McClean2018landscapes}.

In brief, while the Haar measure represents the theoretical upper bound of VQCs' expressivity, it is often unattainable in practice due to excessive computational resource demands. $t$-designs provide a practical alternative to enhance VQCs' expressivity by approximating Haar randomness with significantly reduced costs.

While $t$-designs mitigate the computational challenges associated with Haar randomness, they can still result in highly random unitary operators, particularly as the degree increases. This introduces a critical issue in VQCs' training known as barren plateaus (BPs). This issue poses significant challenges to the practical deployment of VQCs. To generally define the BPs, Qi et al.~\cite{qi2023barren} provide a simple but generic definition as follows:
\begin{equation}
    Var[\partial C] \leq F(N),
\label{eq:bps}
\end{equation}
where $Var[\partial C]$ denotes the variance of gradient, $F(N) \in o\left( \frac{1}{b^N} \right)$, for some $b > 1$ and, $N$ denotes the number of qubits in VQCs.

According to Equation~\ref{eq:bps}, $Var[\partial C]$ will exponentially decrease to zero when the number of qubits $N$ increases. In this case, most gradient-based approaches will fail to train VQCs. Therefore, it is critical to mitigate barren plateau issues to train VQCs robustly.
Notably, in this survey, we aim to investigate the mitigation methods designed to avoid the occurrence of barren plateaus at the beginning of the training process. Most mitigation methods theoretically or empirically verify the effectiveness under certain conditions, such as specific ranges of qubits or layers. The results indicate that the proposed methods can avoid BPs under these circumstances but do not necessarily guarantee the elimination of BPs for all cases. Thus, we use the term ``mitigate" to describe these efforts.
\section{Investigation of Barren Plateaus}
\label{sec:investigations}

In this section, we will introduce several works that aim to investigate barren plateaus (BPs). First, we introduce some works that explore and understand the barren plateau phenomena in Section~\ref{sec:understanding}. Furthermore, we present the related works that investigate BPs from the perspective of model architectures in Section~\ref{sec:understanding1} and optimization in Section~\ref{sec:understanding2}. Last, we discuss several works that provide tools to evaluate the strategies for mitigating BPs in Section~\ref{sec:evaluations}.

\subsection{Understanding Barren Plateau Phenomena}
\label{sec:understanding}

Since the BP phenomena are categorized, several studies continue to delve into whether any other conditions or reasons may trigger the same issue. In this section, we overview related works that offer an understanding of the barren plateaus.

To dive deeper into BPs, Wang et al.~\cite{wang2021noise} further prove that the gradient will vanish exponentially under the consideration of local Pauli noise, which is quite different from the noise-free setting in~\cite{McClean2018landscapes}. Similarly, Singkanipa et al.~\cite{singkanipa2024beyond} explore the effects of noise on VQCs, focusing on how non-unital noise contributes to the formation of barren plateaus and fixed points in the optimization landscape. 
Marrero et al.~\cite{marrero2021entanglement} show that excessive entanglement between the visible and hidden units in VQCs can hinder learning capacity.
Holmes et al.~\cite{holmes2021barren} first investigate the scrambling processes and show that showing that any variational ansatz is highly probable to have a barren plateau landscape.
Later, Holmes et al.~\cite{holmes2022connecting} study the expressibility of VQCs and reveal that highly expressive VQCs may exhibit flatter optimization landscapes. Similarly, Ragone et al.~\cite{ragone2023unified} provide a unified framework to study the expressibility of VQCs.
Liu et al.~\cite{liu2024laziness} theoretically provide an understanding of the distinction between BPs and ``laziness''. In this study, laziness refers to the exponential suppression of variational angle updates during gradient descent. By studying laziness, researchers have a deeper understanding of the ansatz structure as a trade-off between flat landscapes and trainability, i.e., trainability in the presence of BPs via laziness).
Friedrich et al.~\cite{friedrich2024barren} examine how the dimensionality of qudits influences the occurrence of BPs in VQCs.
Fontana et al.~\cite{fontana2024adjoint} introduce the concept of adjoint differentiation to analyze BPs and provide theoretical insights for designing quantum ansätze to avoid barren plateaus. Comparably, Diaz et al.~\cite{diaz2023showcasing} extend the theoretical understanding of BPs beyond the conventional dynamical Lie algebra (DLA) framework and propose new criteria to identify BPs, while Goh et al.~\cite{goh2023lie} leverage Lie-algebraic tools~\cite{patti2021entanglement} to develop a classical simulation method, g-sim, that efficiently simulates BP-free unitary circuits. Last, Anschuetz et al.~\cite{anschuetz2022quantum} investigate other untrainable scenarios beyond barren plateaus.

\subsection{Investigations from Model Perspectives}
\label{sec:understanding1}
A portion of the research works aims to investigate BPs from the perspective of existing quantum circuits.
Wiersema et al.~\cite{wiersema2020exploring} first examine a family of VQCs, named the Hamiltonian Variational Ansatz (HVA), and reveal that HVA is absent from BPs due to its favorable structural properties. Mao et al.~\cite{mao2023barren} theoretically investigate the existence of Barren Plateaus for alternated disentangled UCC (dUCC) ansatz.
Besides, Liu et al.~\cite{liu2022presence} provide critical insights into the trainability of tensor-network models and reveal that barren plateaus are absent near the local minimum. This discovery has led to many investigations in new VQC architectures. For example, Martin et al.~\cite{martin2023barren} investigate the BP issues in several randomly chosen parameterized circuits and reveal that some model architectures like tree tensor network (qTTN) and multi-scale entanglement re-normalization ansatz (qMERA) could avoid BPs. Cybulski et al.~\cite{cybulski2023impact} investigate four scenarios about barren plateaus by examining the model depth, layer-by-layer pre-training, circuit block structure, and model creation without any constraints.
Beyond evaluating VQCs, several works validate that BPs don't exhibit in classical-quantum hybrid models. Abbas et al.~\cite{abbas2021power} empirically demonstrate that a class of quantum neural networks (QNNs) can outperform comparable feedforward networks and prove that certain QNNs can avoid BPs because of the unique optimization landscapes. Similarly, Pesah et al.~\cite{pesah2021absence} analyze the cost function of QCNNs and show that BPs don't exhibit in QCNNs, indicating the potential of QCNNs for advancing NISQ computing. Coelho et al.~\cite{coelho2024vqc} explore the trainability and performance of VQCs on deep Q-learning models in classical environments and indicate that increasing the number of qubits does not lead to BPs for these models.

\subsection{Investigations from Optimization Perspectives}
\label{sec:understanding2}
Another portion of the research works aims to investigate BPs from the perspective of the optimization process.
Zhao et al.~\cite{zhao2021analyzing} propose a scheme to analyze the barren plateau phenomenon in training VQCs with the ZX-calculus. Cerezo et al.~\cite{cerezo2021cost} study the connections between locality and trainability of different cost functions in shallow VQCs and reveal that optimizing VQCs via cost functions with global observables exhibit BPs, whereas the gradient vanishes polynomially when using local observables. Concurrently, Cerezo et al.~\cite{cerezo2021higher} verify that Hessian-based approaches do not circumvent the exponential scaling associated with BPs. Later, Cerezo et al.~\cite{cerezo2023does} also reveal the cost function can be classically simulated when the gradient variance vanishes. This result arises from a curse of dimensionality. Current methods essentially encode this problem into small and classically simulatable subspaces.
Arrasmith et al.~\cite{arrasmith2021effect} empirically verify that the gradient-free optimizer does not solve the barren plateau problem. Furthermore, Arrasmith et al.~\cite{arrasmith2022equivalence} investigate the interdependent connections among three optimization landscapes, barren plateaus, exponential cost concentration around the gradient mean, and narrow gorges, during the training phase of VQCs. This work indicates that in high-dimensional space, BPs will cause cost concentration and thus reduce the training efficiency. Similarly, Miao et al.~\cite{miao2024equivalence} evaluate BP issues and show that BPs would be addressed in the Riemannian formulation of such optimization problems. Furthermore, Miao et al.~\cite{barthel2023absence, miao2024isometric} provide an in-depth analysis of isometric tensor network optimization methods applied to extensive Hamiltonians, i.e., VQCs with extensive size, and demonstrate that these approaches are inherently free of barren plateaus. Also, these works evaluate the gradient variance of tensor-network-based approaches and their scaling property. Cao et al.~\cite{cao2024exploiting} explore how many-body localization (MBL)-thermalization phase transitions can be utilized to avoid BPs, indicating that MBL states can be leveraged to optimize the design of quantum circuits. Pérez-Salinas et al.~\cite{perez2024analyzing} investigate landscapes that contain relevant information causing optimization hardness, contributing connections between information content and the scale of the gradient. Nemkov et al.~\cite{nemkov2024barren} investigate BPs in the presence of local minima and reveal that enlarging gradients cannot sufficiently address BPs under the above circumstance.
From the theoretical perspectives in optimization, Uvarov et al.~\cite{uvarov2021barren} provide a lower bound for training VQCs and argue that the magnitude of gradient variance can be heavily influenced by the Hamiltonian and circuit structure. Letcher et al.~\cite{letcher2023tight} study the bounds of gradient descent in VQCs, providing tight upper and lower bounds during VQCs' training for different amounts of observables.

\subsection{Evaluations}
\label{sec:evaluations}
Several works provide frameworks or tools to evaluate the strategies that can mitigate BPs. In this section, we discuss several tools as follows. Patti et al.~\cite{patti2021entanglement} provide a toolbox that integrates several techniques, such as initialization or regularization strategies, to mitigate BPs. This toolbox enables developers to experiment and tinker around with additional entanglement and noise in the circuit. Similarly, Larocca et al.~\cite{larocca2022diagnosing} introduce a theoretical framework to diagnose the presence of BPs for problem-based ansatz such as QAOA and HVA. This research focuses on utilizing concepts from quantum optimal control (i.e. dynamical lie algebra) to predict whether BPs will arise in the training landscape of problem-based ansatz. Park et al.~\cite{park2024quantum} introduce a new software engineering tool that aims at assisting developers in efficiently writing, debugging, and testing code related to BP experiments. This tool is a run-time testing, analysis, and code optimization (TACO) for QNNs in advanced IoT systems. Kashif et al.~\cite{kashif2024alleviating} analyze the impact of initialization-based strategies from classical machine learning in random VQCs from the aspect of the BP phenomenon.
In brief, we summarize the keywords for these toolboxes, which can help researchers write code and conduct barren plateau experiments, in Table~\ref{tab:toolTable}.
\begin{table}[h]
\centering
\scriptsize
\caption{Summary of keywords in the toolbox for mitigating BPs.}
\label{tab:toolTable}
\begin{tabular}{cc}
\toprule
{\bf Authors} & {\bf Keywords} \\
\midrule
Patti et al.~\cite{patti2021entanglement} &  Entanglement metrics, Meta-learning, Solution Factorization \\
Larocca et al.~\cite{larocca2022diagnosing} &  Quantum control, dynamical lie algebra, BP Diagnosis \\
Park et al.~\cite{park2024quantum} & Code optimization, gradient visualization, QNN IoT systems \\
Kashif et al.~\cite{kashif2024alleviating} & Evaluate initialization-based strategies \\
\bottomrule
\end{tabular}
\end{table}

\section{Mitigation of Barren Plateaus}
\label{sec:mitigations}

In this section, we categorize existing mitigation strategies into five groups from the view of initialization, optimization, model architecture, regularization, and measurement. Besides, we summarize and discuss the strategies for each group.

\subsection{Initialization-based Strategies}
\label{subsec:initialization}
\begin{figure}[h!]
    \centering
    \includegraphics[scale=0.7]{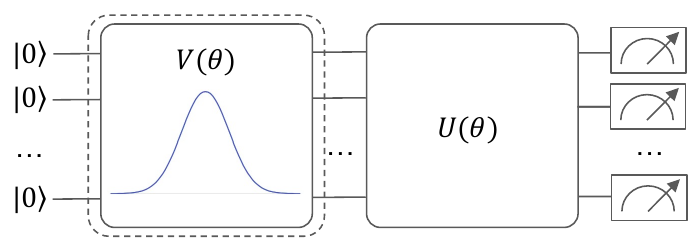}
    \caption{A general structure of a VQC where the initialization-based strategy is circled in a dotted box. Generally speaking, a VQC will take the encoded quantum-state data as inputs and learn the data representation through a sequence of unitary operations. For each unitary gate, VQC has a corresponding parameter. In the beginning, initialization-based strategies $V(\theta)$ are applied to initialize parameters $\theta$ of the unitary quantum circuits $U(\theta)$ before training. In the end, a measurement layer is applied to measure the output of the VQC.}
    \label{fig:initialization}
\end{figure}

In this section, we explore various initialization-based strategies applied to VQCs. Within this category, most strategies aim to employ different initialization methods to help VQCs get rid of barren plateaus in the initialization stage. To better illustrate these strategies, we present a general idea in Figure~\ref{fig:initialization}.
In the following paragraph, we will briefly introduce each work within this category.

Grant et al.~\cite{grant2019initialization} propose a strategy that first randomly initializes some gates and further sets the rest gates to produce the identity matrix, ensuring initial gradients non-vanishing.
Sauvage et al.~\cite{sauvage2021flip} introduce a flexible initialization strategy that can adapt suitable initial parameters to various sizes of quantum circuits.
Sack et al.~\cite{sack2022avoiding} design a method to circumvent weak barren plateaus by using classical shadows protocol at the initialization stage and during the training process.
Rad et al.~\cite{rad2022surviving} introduce a fast-and-slow algorithm that employs the Bayesian Learning approach to initialize parameters for VQCs by initializing a fast local optimizer to find the global optimum point efficiently. In this algorithm, the Bayesian approach can help predict effective starting points, thereby reducing the risk of barren plateaus.
Kulshrestha et al.~\cite{kulshrestha2022beinit} first initialize the model parameters using a beta distribution, which can better model the data distribution in binary classification, and further add noise into the parameters during training to avoid the optimization being trapped in saddle points. They empirically verify that utilizing beta distribution can effectively mitigate the rapid decrease of gradient variance in initialization as the model size increases.
Similarly, Zhang et al.~\cite{zhang2022escaping} employ a Gaussian distribution as initial model parameters and demonstrate that such a strategy can effectively help VQCs avoid BPs.
Friedrich et al.~\cite{friedrich2022avoiding} utilize a classical neural network, namely a convolutional neural network (CNN), to randomly generate the model parameters, indicating the effectiveness of this method for mitigating BPs.
Mele et al.~\cite{mele2022avoiding} offer an iterative search scheme to transfer smooth solutions from smaller systems as a warm start for larger systems. Such a scheme has been proven to help optimization avoid the barren plateau issue.
Grimsley et al.~\cite{grimsley2023adaptive} first examine how the Adaptive Problem-Tailored Variational Quantum Eiegensolver (ADAPT-VQE) ansätze is affected by local minima and further propose a strategy that can initialize parameters with over an order of magnitude smaller error than random initialization.
Liu et al.~\cite{liu2023mitigating} present a transfer-learning approach where model parameters learned from small-scale tasks are transferred to larger tasks. This method leverages the smooth solutions from simpler problems to provide a good starting point for more complex VQCs, therefore avoiding barren plateaus and enhancing training efficiency.
Park et al.~\cite{park2024hamiltonian} first show that a time-evolution operator generated by a $k$-local Hamiltonian can help VQCs avoid BPs and then propose an initialization scheme using this operator. The $k$-local Hamiltonian can perform efficiently as it acts non-trivially on at most $k$ qubits~\cite{kempe2006complexity, cubitt2016complexity}, such as addressing the contains with finite correlation length in quantum spin lattice systems~\cite{anshu2016concentration}. Some studies discussed in this survey also utilize the concept of $k$-local Hamiltonian~\cite{sack2022avoiding, park2024hardware}.
Overall, initialization-based strategies can be summarized in Table~\ref{tab:initialization}.
\begin{table}[h]
\centering
\small
\caption{Summary of initialization-based strategies.}
\label{tab:initialization}
\begin{tabular}{cc}
\toprule
{\bf Authors} & {\bf Methods} \\
\midrule
Grant et al.~\cite{grant2019initialization} & Identity block strategy \\
Sauvage et al.~\cite{sauvage2021flip} & Flexible strategy \\
Sack et al.~\cite{sack2022avoiding} & Classical shadows protocol \\
Rad et al.~\cite{rad2022surviving} & Fast-and-slow algorithm \\
Kulshrestha et al.~\cite{kulshrestha2022beinit} & Beta initialization \\
Zhang et al.~\cite{zhang2022escaping} & Gaussian initialization \\
Friedrich et al.~\cite{friedrich2022avoiding} & Initialization via CNNs \\
Mele et al.~\cite{mele2022avoiding} & Transfer from small to large \\
Grimsley et al.~\cite{grimsley2023adaptive} & ADAPT-VQE \\
Liu et al.~\cite{liu2023mitigating} & Transfer-learning method \\
Park et al.~\cite{park2024hamiltonian} & A time-evolution operator \\
\bottomrule
\end{tabular}
\end{table}

\subsection{Optimization-based Strategies}
\label{subsec:optimization}
\begin{figure}[h]
    \centering
    \includegraphics[scale=0.7]{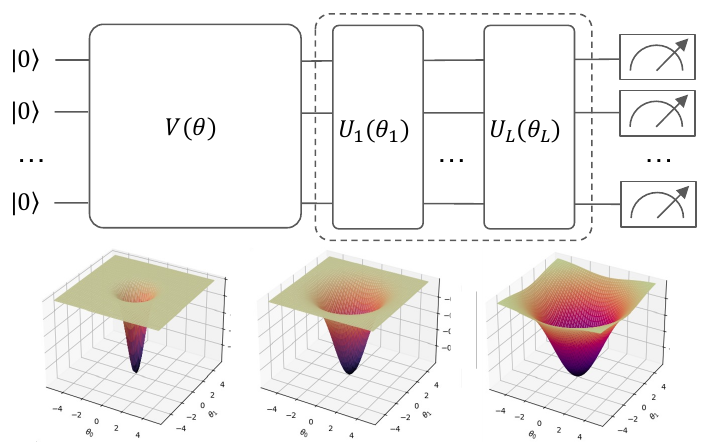}
    \caption{A general structure of VQCs where the optimization-based strategies are usually applied in the training of $U(\theta)$ (dotted box), which is decomposed into $L$ layer from $U_1(\theta_1)$ to $U_L(\theta_L)$ (upper side). On the lower side, we present the process for addressing BPs. The left-most cost-function landscape represents BPs, a flat landscape with no discernible slope towards the minimum. In this case, many optimization approaches may be initially trapped in the flat landscape, leading to a failure of training. By employing optimization-based strategies, the cost-function landscape could be gradually recovered, as shown in the middle and right-most landscapes.}
    \label{fig:optimization}
\end{figure}

In this section, we go over the optimization-based strategies that mainly claim to address BPs in optimization, which is an essential step for optimizing VQCs~\cite{nadori2024promising}. To better visualize the optimization-based strategies, we present a general layout in Figure~\ref{fig:optimization}. Note that many optimization-based strategies can enhance the trainability of VQCs but some of them only solve the gradient vanishing problem during training rather than focusing on BPs. To follow the topic of this survey, we don't discuss these papers in this section.

Ostazewski et al.~\cite{ostaszewski2021structure} propose a gradient-free algorithm, {\it Rotoselect}, to enhance the optimization process of VQCs by sequentially adjusting angles of rotation and circuit structure to efficiently minimize the energy. This algorithm can significantly enhance the performance of shallow circuits compared to those relying solely on parameter updates, rendering this method particularly advantageous for noisy intermediate-scale quantum computers.
On the contrary, most works alleviate BPs via gradient-based approaches.
Skolik et al.~\cite{skolik2021layerwise} propose a layer-wise training strategy that can incrementally grow the depth of VQCs during optimization and only update a subset of model parameters in each training epoch. This work also verifies the effectiveness of addressing BPs.
Similarly, Gharibyan et al.~\cite{gharibyan2023hierarchical} introduce a new hierarchical framework that can incrementally learn at different scales of VQCs. This framework starts from a block with a small number of qubits and incrementally expands the blocks with a larger number of qubits. Such a hierarchical design can not only ensure the performance of VQCs but also avoid BPs.
The above two methods commonly grow the model size during training, but the main difference between them is that the layer-wise learning method~\cite{skolik2021layerwise} stacks single-rotation gates and two-qubit gates to form all-to-all connected layers successively, whereas the hierarchical learning method~\cite{gharibyan2023hierarchical} expands the blocks with a larger number of qubits.
From the perspective of parameters reduction, compared to layer-wise learning~\cite{skolik2021layerwise}, which updates a subset of parameters iteratively, Liu et al.~\cite{liu2024mitigating} propose a state-efficient ansatz (SEA) to accurately estimate the ground state with much fewer parameters than common VQCs while efficiently mitigating barren plateaus and improving the trainability of VQCs.

Furthermore, several approaches attempt to mitigate BPs via new proposed mechanisms in optimization.
Tobias Haug and M. S. Kim~\cite{haug2021optimal}, propose methods to enhance the training process of VQCs by leveraging Gaussian kernels to derive adaptive learning rates during optimization and by introducing the generalized quantum natural gradient. The proposed methods ensure the stability of training and reduce the risk of barren plateaus.
Kieferova et al.~\cite{kieferova2021quantum} examine the assumptions of BPs and indicate that the issues can be circumvented by an unbounded loss function. Based on this examination, they propose a training algorithm that minimizes the maximal Rényi divergence of order two. They further verify the effectiveness of the proposed algorithm on quantum neural networks (QNNs) and quantum Boltzmann machines (QBM) across two cases, thermal state learning and Hamiltonian learning.
Heyraud et al.~\cite{heyraud2023efficient} provide an efficient method to compute the gradient of the cost function and its variance for a wide class of VQCs by mapping from randomly initialized circuits to a set of Clifford circuits, which can be simulated on classical devices. The method can not only mitigate BPs but also increase the scalability of VQCs.
Sannia et al.~\cite{sannia2023engineered} aim to mitigate barren plateaus and enhance the trainability of VQCs via engineered Markovian dissipation processes. To achieve this goal, the proposed method artificially manufactures additional noise after each gate in the VQC. The experiments show that the added noise can help restore the trainability of a VQC by mitigating the exponential scaling of the gradient's variance.
Mele et al.~\cite{mele2024noise} examine the role of noise in creating shallow quantum circuits that do not exhibit barren plateaus for cost functions composed of local observables.
Broers et al.~\cite{broers2024mitigated} propose an optimization approach based on trainable Fourier coefficients of Hamiltonian system parameters and demonstrate that this approach can mitigate BPs.
Zambrano et al.~\cite{zambrano2024avoiding} address BP issues in the variational quantum computation of geometric entanglement and present an approach to maintain gradients throughout the optimization process of VQCs.

Besides improving the trainability of VQCs, another group of optimization-based methods addresses BPs from the perspective of data encoding. We categorize them in this section as optimization is the main contribution of these methods while their proposed encoding mechanisms can address BPs.
Sciorilli et al.~\cite{sciorilli2024towards} introduce a hybrid
quantum-classic solver for combinatorial optimizations and further analyze that employing specific qubit-efficient encoding methods can mitigate barren plateaus. Also, this work verifies the effectiveness of the proposed methods on max-cut problems.
Falla et al.~\cite{falla2024graph} first investigate several graph embedding techniques for parameter transferability, such as the transferability between different classes of max-cut instances, and further effectively mitigate BPs during variational optimization. Although this is a naive approach, it effectively reduces the number of iterations in optimization, obtaining a speedup for parameter convergence.

Overall, the above optimization-based strategies are summarized in Table~\ref{tab:optimization}.
\begin{table}[h]
\centering
\scriptsize
\caption{Summary of optimization-based strategies.}
\label{tab:optimization}
\begin{tabular}{cc}
\toprule
{\bf Authors} & {\bf Methods} \\
\midrule
Ostaszewski et al.~\cite{ostaszewski2021structure} & A gradient-free algorithm, ``Rotoselect'' \\
Skolik et al.~\cite{skolik2021layerwise} & Layer-wise learning method \\
Gharibyan et al.~\cite{gharibyan2023hierarchical} & Hierarchical variational circuit framework \\
Liu et al.~\cite{liu2024mitigating} & State-efficient ansatz \\
Haug et al.~\cite{haug2021optimal} & Adaptively adjust learning rates \\
Kieferova et al.~\cite{kieferova2021quantum} & Rényi 2-divergence \\
Heyraud et al.~\cite{heyraud2023efficient} & Map randomly circuits to Clifford circuits \\
Sannia et al.~\cite{sannia2023engineered} & Engineered Markovian dissipation processes \\
Mele et al.~\cite{mele2024noise} & Noise-induced shallow circuits \\
Broers et al.~\cite{broers2024mitigated} & Introduce a Fourier-based ansatz \\ 
Zambrano et al.~\cite{zambrano2024avoiding} & Evaluate geometric entanglement \\
Sciorilli et al.~\cite{sciorilli2024towards} & Pauli-correlations encoding \\
Falla et al.~\cite{falla2024graph} & Graph embedding \\
\bottomrule
\end{tabular}
\end{table}

\subsection{Model-based Strategies}
\label{subsec:architecture}
\begin{figure}[h]
    \centering
    \includegraphics[scale=0.8]{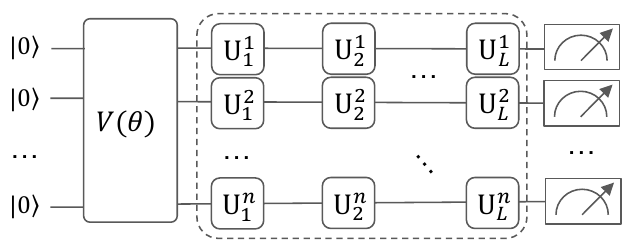}
    \caption{An overall idea of model-based strategies that are usually applied to the decomposed unitary circuit $U(\theta) \in \mathbf{C}^{n\cdot L}$ (dotted box), where $U_L^n$ denotes that $U(\theta)$ consists of $n$ qubits and $L$ layers.}
    \label{fig:model}
\end{figure}

In this section, we explore several model-based strategies that propose new model architectures to address BP issues over the past few years~\cite{yao2024avoiding, friedrich2024quantum, park2024hardware, shi2024avoiding}. These methods encompass a wide variety of new circuit structures and new gradient-based or gradient-free frameworks. An overall idea of model-based strategies is presented in Figure~\ref{fig:model}.

Within this category, Li et al.~\cite{li2021vsql} first propose a new hybrid quantum-classical framework, Variational Shadow Quantum Learning (VSQL), that utilizes the classical shadows of quantum data, a kind of side information of quantum data, in supervised learning. This framework can not only reduce noise but also significantly avoid BP issues.
Bharti et al.~\cite{bharti2021quantum} pioneer a groundbreaking approach to quantum simulation by introducing a novel quantum-classical algorithm. Unlike most existing methods, their algorithm operates without relying on any classical-quantum feedback loop during training. Notably, their work demonstrates to be free from BP problems, marking a significant advancement in simulating the dynamics of quantum systems.
Du et al.~\cite{du2022quantum} introduce a new search scheme called quantum architecture search (QAS), aimed at identifying approximately optimal ansatz structures by assessing the expressivity of VQCs. This innovative approach enables QAS to dynamically generate ansatz structures tailored to specific tasks, offering a flexible solution to address challenges such as mitigating the negative effects of quantum noise and bypassing BP issues.
Zhang et al.~\cite{zhang2022quark} introduce a gradient-free framework, Quark, that optimizes the models via quantum optimization. This framework can avoid BPs and frequent classical-quantum interactions because it doesn't compute the gradient during model optimization.
Selvarajan et al.~\cite{selvarajan2023dimensionality} design a variational-autoencoder-based circuit to reduce states represented in large Hilbert spaces with half the number of qubits while retaining the features of the starting state. This reduction can potentially help improve training efficiency on higher dimensional Hilbert spaces and thus overcome BP issues.
Tuysuz et al.~\cite{tuysuz2023classical} propose a classical splitting ansatz that decomposes an n-qubit circuit into circuits of size $O(\log N)$. This ansatz has been shown to mitigate BPs with the demonstration of their absence in tasks such as binary classification.
Inspired by classical residual neural networks, Kashif et al.~\cite{kashif2024resqnets} introduce residual connections in VQC. This proposed model, a.k.a., resQNets, breaks down a VQC structure into smaller nodes and introduces residual connections between these nodes. The experimental results show that resQNets can outperform conventional VQCs and offer a new model architecture to solve BP issues.
Shin et al.~\cite{shin2024layerwise} propose a unified methodology, layerwise quantum convolutional neural networks (LQCNNs), to avoid BPs and to provide a solution for large qubit states. This study rigorously proves that LQCNNs preserve fundamental quantum properties, such as different entropies and fidelity of quantum states, despite their deep structure.
Zhang et al.~\cite{zhang2024absence} theoretically analyze lower bounds on gradient variances of VQCs and propose a unified framework, finite local-depth circuits (FLDC), to mitigate BPs.
In brief, we summarize the model-based strategies in Table~\ref{tab:model}.
\begin{table}[h]
\centering
\scriptsize
\caption{Summary of model-based strategies.}
\label{tab:model}
\begin{tabular}{cc}
\toprule
{\bf Authors} & {\bf Methods} \\
\midrule
Li et al.~\cite{li2021vsql} & Utilizes side information of quantum data \\
Bharti et al.~\cite{bharti2021quantum} & Training without a feedback loop \\
Du et al.~\cite{du2022quantum} & Dynamically generate ansatz structures \\
Zhang et al.~\cite{zhang2022quark} & Introduce a gradient-free framework \\
Selvarajan et al.~\cite{selvarajan2023dimensionality} & Design a variational-autoencoder-based circuit \\
Tuysuz et al.~\cite{tuysuz2023classical} & Propose a splitting ansatz to decompose circuit \\
Kashif et al.~\cite{kashif2024resqnets} & Introduce residual connections in VQC \\
Shin et al.~\cite{shin2024layerwise} & Propose a unified layerwise QNN \\
Zhang et al.~\cite{zhang2024absence} & Propose a unified framework, FLDC \\
\bottomrule
\end{tabular}
\end{table}

\subsection{Regularization-based Strategies}
\label{sec:regularization}
Regularization is a machine-learning technique to constrain the search space for boosting optimization performance. In other words, regularization achieves its goal by penalizing complex model parameters $\theta$, effectively generalizing the model to unseen data. In general, regularization can be formulated as follows.
\begin{equation}
  \min_{\theta} [C(\theta)+\lambda Reg(\theta)],
\label{eqn:reg}
\end{equation}
where $Reg(\cdot)$ denotes the regularization method and $\lambda$ denotes its coefficient.

\begin{figure}[h]
    \centering
    \includegraphics[scale=0.7]{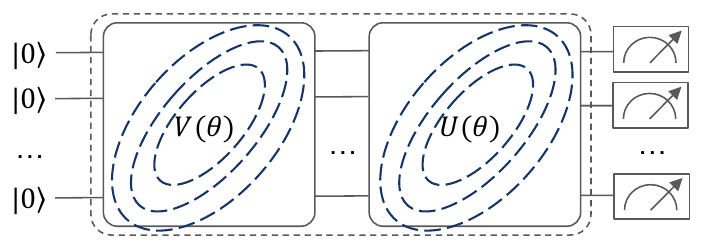}
    \caption{An overall idea of regularization-based strategies. This strategy could be applied to the model parameters during initialization or training. We use dotted circles to represent the regularization in this figure.}
    \label{fig:regularization}
\end{figure}

In this survey, regularization-based strategies aim to apply regularization to VQCs. The regularization could be applied to the model parameters in the initialization stage or the optimization stage. We present the overall idea of this strategy in Figure~\ref{fig:regularization}.

Patti et al.~\cite{patti2021entanglement} utilize regularization to penalize entanglement and show that this method could ameliorate BPs and further decrease both training time and error. Zhuang et al.~\cite{zhuang2024improving} propose regularization methods that can be applied during initialization and optimization. Specifically, the proposed methods regularize the initial distribution of model parameters with prior knowledge of the train data and iteratively add diffused Gaussian noise to the parameters during training. This is the first work that leverages regularization strategies to simultaneously address barren plateaus and saddle points in optimizing VQCs.

\subsection{Measurement-based Strategies}
\label{sec:measurement}
\begin{figure} [h]
    \centering
    \includegraphics[scale=0.95]{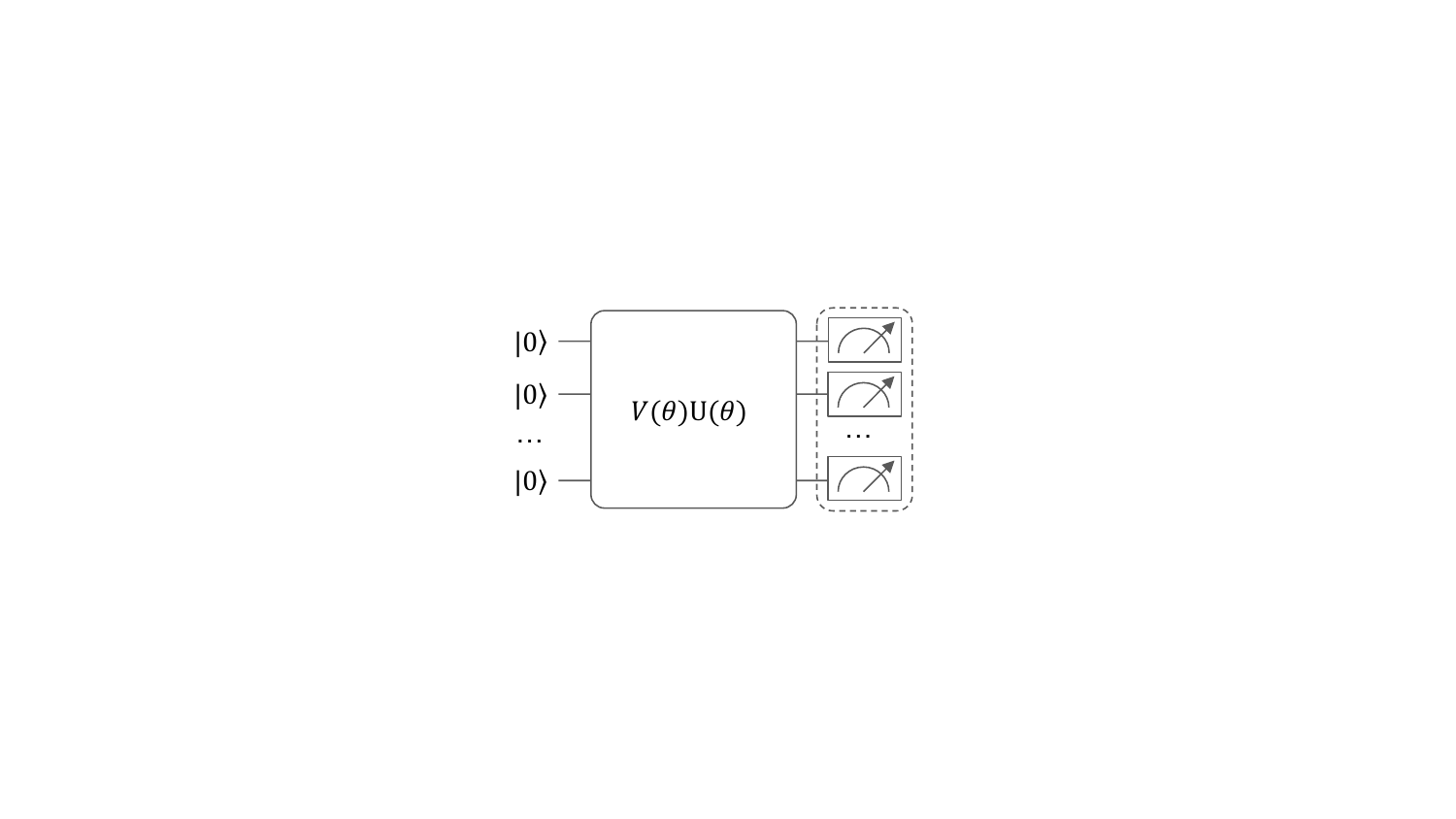}
    \caption{An example of measurement-based strategy, which is applied to the measurement of the VQC $V(\theta)U(\theta)$, where $V(\theta)$ denotes the initialization module and $U(\theta)$ denotes the unitary circuit. The measurement step (shown within the dotted box) typically focuses on measuring the output of the VQCs.}
    \label{fig:measurement}
\end{figure}

Measurement techniques aim to provide more informative metrics about the qubits in the system during and after the optimization process. By obtaining richer information about the quantum data, researchers may gain a deeper understanding of what their data is and structure the VQC with the data accordingly. This allows for more efficient training and higher performance in VQCs. In Figure~\ref{fig:measurement}, we show an example of a measurement-based strategy.

One measurement-based strategy introduced by Rappaport et al.~\cite{rappaport2023measurement} demonstrates that entanglement-induced BPs could be suppressed by adding measurements with post-selection. This work empirically studies two measures, measurement-induced landscape transition (MILT) and measurement-induced phase transition (MIPT). and further reveals the benefits of using MILT for optimization. 

\begin{table*}[t]
\centering
\caption{Summary of mitigation strategies for barren plateau issues. ``$\| N \|$" and ``$\| L \|$" denote the maximum number of qubits and layers in the corresponding paper. ``Problem Type" denotes the problem that the work mainly focuses on.}
\label{tab:sum_mitigation}
\begin{tabular}{c|c|c|c|c}
    \toprule
    Authors & Data Encoding & $\| N \|$ & $\| L \|$ & Problem Type\\
    \midrule
    \multicolumn{5}{c}{Initialization-based Strategies}\\
    \hline
    Grant et al.~\cite{grant2019initialization} & Amplitude & 12 & 120 & Ground State \\
    Sauvage et al.~\cite{sauvage2021flip} & FLIP Encoding-decoding & 16 & 16 & MaxCut, Ground State \\
    Sack et al.~\cite{sack2022avoiding} & $\times$ & 16 & 100 & Ground state \\
    Rad et al.~\cite{rad2022surviving} & $\times$ & 6 & 2 & Algorithm evaluation \\
    Kulshrestha et al.~\cite{kulshrestha2022beinit} & Phase & 10 & 30 & Classification \\
    Zhang et al.~\cite{zhang2022escaping} & Basis & 15 & 72 & Ground State \\
    Friedrich et al.~\cite{friedrich2022avoiding} & Qubit encoding  & 10 & 30 & Model evaluation \\    
    Mele et al.~\cite{mele2022avoiding} & Symmetry encoding & 20 & 100 & Ground State \\
    Grimsley et al.~\cite{grimsley2023adaptive} & $\times$ & 14 & $\times$ & Ground State \\    
    Liu et al.~\cite{liu2023mitigating} & Block identity encoding  & 4 & 4 & Ground State \\
    Park et al.~\cite{park2024hamiltonian} & Local Hamiltonian evolution & 28 & $\times$ & Thermalization \\
    \hline
    \multicolumn{5}{c}{Optimization-based Strategies}\\
    \hline
    Ostaszewski et al.~\cite{ostaszewski2021structure} & $\times$ & 7 & 15 & Ground State \\
    Skolik et al.~\cite{skolik2021layerwise} & Phase & 18 & 200 & Classification \\
    Haug et al.~\cite{haug2021optimal} & $\times$ & 25 & 20 & Ground state \\
    Kieferova et al.~\cite{kieferova2021quantum}& $\times$  & $\times$ & $\times$ & Assessment of generative training \\
    Heyraud et al.~\cite{heyraud2023efficient} & $\times$ & 100 & 10 & Ground State \\
    Gharibyan et al.~\cite{gharibyan2023hierarchical} & Amplitude & 27 & 9 & Assessment of QCBM models \\
    Sannia et al.~\cite{sannia2023engineered} & $\times$ & 10 & 20 & Ground State \\ 
    Mele et al.~\cite{mele2024noise} & $\times$ & 6 & 10 & $\times$ \\
    Broers et al.~\cite{broers2024mitigated} & $\times$ & 8 & $\times$ & Ground State \\
    Zambrano et al.~\cite{zambrano2024avoiding} & $\times$ & 18 & $\times$ & Ground State \\
    Liu et al.~\cite{liu2024mitigating} & $\times$ & 18 & 2 & Ground state \\
    Sciorilli et al.~\cite{sciorilli2024towards} & Pauli-correlations  & 17 & 11 & MaxCut \\
    Falla et al.~\cite{falla2024graph} & Graph embedding & 27 & 3 & MaxCut \\
    \hline
    \multicolumn{5}{c}{Model-based Strategies}\\
    \hline
    Li et al.~\cite{li2021vsql} & Phase & 10 & 5 & Classification \\
    Bharti et al.~\cite{bharti2021quantum} & $\times$ & 12 & 200 & Ground State \\
    Du et al.~\cite{du2022quantum} & Phase & 10 & 3 & Ground State, Classification \\
    Zhang et al.~\cite{zhang2022quark} & Basis & 4 & 2 & Edge Detection, Classification \\
    Selvarajan et al.~\cite{selvarajan2023dimensionality} & Amplitude & 8 & 5 & Classification \\
    Tuysuz et al.~\cite{tuysuz2023classical} & Phase  & 24 & 200 & Classification \\
    Kashif et al.~\cite{kashif2024resqnets} & $\times$ & 20 & 6 & Analyze cost function landscapes \\
    Shin et al.~\cite{shin2024layerwise} & $\times$ & $\times$ & $\times$ & $\times$ \\
    Zhang et al.~\cite{zhang2024absence} & $\times$ & 12 & 16 & Ground State \\
    \hline
    \multicolumn{5}{c}{Regularization-based Strategies}\\
    \hline
    Patti et al.~\cite{patti2021entanglement} & $\times$ & 9 & 200 & Ground State \\
    Zhuang et al.~\cite{zhuang2024improving} & Phase & 16 & 50 & Classification \\
    \hline
    \multicolumn{5}{c}{Measurement-based Strategies}\\
    \hline
    Rappaport et al.~\cite{rappaport2023measurement} & $\times$  & 18 & 100 & Ground State \\
    \hline
\end{tabular}
\end{table*}

\section{Discussion}
\label{sec:discussion}

In this section, we first present the summary of mitigation strategies in a table. Second, we compare our work with the concurrent survey papers. Third, we highlight the similarity and distinction between gradient issues and barren plateaus. Last, we provide insightful discussion on the future directions.

\subsection{Summary of Mitigation Strategies}
In Table~\ref{tab:sum_mitigation}, we summarize several types of mitigation methods for barren plateaus. The category is proposed from the perspective of initialization, optimization, model architecture, regularization, and measurement. In the title of the table, the attributes ``$\| N \|$" and ``$\| L \|$" denote the maximum number of qubits and layers in the corresponding paper. ``Problem Type" denotes the problem that the work mainly focuses on. Note that in most works, authors will examine different numbers of qubits or layers, whereas we only present the maximum numbers mentioned in the paper.
In this table, we have several observations as follows.
First, most existing methods aim to mitigate barren plateaus from the perspective of initialization, optimization, and model architecture, whereas regularization and measurement are two relatively new directions for mitigation.
Second, in the initialization-base strategies, many works examine new data encoding methods, indicating that data encoding is a crucial component for mitigation.
Third, Ground State is a vital problem in physics so many researchers are interested in it.
Fourth, classification is an important task for the application of VQCs, whereas most works use Phase encoding in this task.

\subsection{Comparison with Concurrent Survey Papers about BPs}
In this subsection, we compare three concurrent survey papers about BPs and discuss their pros and cons. Also, we summarize their key contributions in Table~\ref{tab:surveys}.

Qi et al.~\cite{qi2023barren} provide an analysis of the barren plateau phenomenon, which significantly affects the training efficiency of quantum neural networks (QNNs). The review categorizes various causes of barren plateaus, including cost function design, circuit depth, and parameter initialization. It also discusses current strategies to mitigate this issue, such as circuit architecture modifications, and highlights emerging trends aimed at overcoming these challenges to improve the scalability and performance of QNNs. This survey is the first publication that introduces related works about BP issues in QNNs but it only mentions limited related work.

Larocca et al.~\cite{larocca2024review} conduct an in-depth review of barren plateaus in variational quantum algorithms (VQAs). This review discusses how various factors such as ansatz design, initial states, measurement observables, loss functions, and hardware noise contribute to the occurrence of barren plateaus. It also highlights current theoretical and heuristic strategies to mitigate these issues. The study of BPs is essential for improving the trainability of VQAs and has implications for related fields like quantum optimal control, tensor networks, and learning theory. This work provides a comprehensive review and larger scope of BPs, including the trainability of VQAs and other gradient issues. However, it doesn't introduce the mitigation strategies in detail.

Gelman et al.~\cite{gelman2024survey} explore various quantum circuits and types of BPs that hinder the optimization of parameterized quantum circuits (PQCs). Specifically, this survey covers the basic concept of PQCs and further describes BPs. It also introduces various types of quantum ansatz and metrics. Furthermore, this paper discusses several reasons that may cause BPs and corresponding mitigation methods. Basically, this work focuses more on the reasons and types of BPs rather than the mitigation methods.

Compared with the above three concurrent survey papers, our work provides a new taxonomy to categorize the existing mitigation strategies for BPs and introduce them in detail. We summarize the above three works in Table~\ref{tab:surveys} to highlight their key contributions.
\begin{table}[h]
\centering
\caption{Comparison with concurrent survey papers about BPs.}
\label{tab:surveys}
\begin{tabular}{p{2.1cm}p{5.5cm}}
\toprule
{\bf Authors} & {\bf Key Contributions} \\
\midrule
Qi et al.~\cite{qi2023barren} & Introduce the origin, solutions, and trends of BPs in QNNs. \\
Larocca et al.~\cite{larocca2024review} & Introduce the frameworks of VQAs and broader topics about BPs. \\
Gelman et al.~\cite{gelman2024survey} & Introduce various types of PQCs and discuss the types of BPs with corresponding mitigation methods. \\
\bottomrule
\end{tabular}
\end{table}

\subsection{Gradient Issues v.s. Barren Plateaus}
Gradient issues are commonly seen during the optimization process of utilizing gradient-based methods. In classical models, common gradient issues include gradient vanishing, local minima, saddle points, etc. Gradient vanishing occurs when the gradient diminishes as it backpropagates through the layers of a neural network, leading to negligible updates in the early layers during training. Local minima refer to points in the loss landscape where the model is stuck in a state that is not the global minimum but cannot escape due to the nature of the gradient descent. Besides local minima, training sometimes gets stuck in a saddle point where the gradient is zero along certain directions, making it difficult for the optimization to proceed efficiently.

In contrast, barren plateaus (BPs) represent a specific type of gradient issue particularly associated with VQCs. Unlike classical gradient issues, BPs occur when the variance of the gradients diminishes rapidly as the number of qubits or layers increases. This rapid decay results in a flat loss landscape, where the gradients are so small that the optimization process becomes effectively stalled right from the beginning of training. Thus, BPs can significantly hinder the training of VQCs.

In this review, we focus on the BP issues occurring in VQCs, examining how this issue arises and its mitigation strategies for the training of VQCs.

\subsection{Future Directions}
Based on the above discussion, in this survey, we present several promising research directions to mitigate barren plateaus in VQCs as follows.
\begin{itemize}
    \item {\bf AI-driven Initialization.} Most initialization-base strategies utilize classical neural networks, or machine learning algorithms to generate initial distribution. In the future, AI-driven methods, such as LLMs, could be a promising tool to adaptively initialize VQC's parameters.
    \item {\bf New Optimization Methods.} Several directions in optimization are worth exploring, such as employing local cost functions to manipulate a smaller subset of qubits to avoid a flat landscape, and improving layer-wise training methods to incrementally avoid BPs in deep VQCs.
    \item {\bf Novel Circuit Architecture.} It is promising to explore new quantum circuit architectures that integrate neural networks to address BPs and further improve the trainability of VQCs.
    \item {\bf Noise-based Techniques.} Noise could induce BPs but also be used for mitigation. For example, noise can be injected into model parameters to avoid a flat landscape. In the meanwhile, these techniques bring us an open question: Will noise-based techniques weaken the optimization of VQCs?
    \item {\bf Graph Embedding.} Besides classic data encoding techniques, embedding data with graph representation could be a potential solution to mitigate BPs.
\end{itemize}
In brief, the above future directions offer researchers new perspectives to address the challenge of barren plateaus in VQCs. These directions pave the effective way for mitigating BPs and further enhance trainability of VQCs.

\section{Conclusion}
\label{sec:con}
In this survey, we conduct a comprehensive overview of related literature about the barren plateaus (BPs) in variational quantum circuits (VQCs) and propose a new taxonomy from the perspective of investigation and mitigation. Especially, we categorize most existing mitigation strategies into five groups from the point of initialization, optimization, model architecture, regularization, and measurement. We introduce these strategies in detail and summarize them in a table. Furthermore, in the discussion section, we compare our work with concurrent survey papers and highlight the similarity and distinction between gradient issues and barren plateaus. Last, we provide insightful discussion on the future directions. We hope that our survey paper can provide more inspiration for researchers who are interested in BP issues.



\small
\bibliographystyle{IEEEtran}  
\bibliography{2reference}

\begin{thebibliography}{10}
\providecommand{\url}[1]{#1}
\csname url@samestyle\endcsname
\providecommand{\newblock}{\relax}
\providecommand{\bibinfo}[2]{#2}
\providecommand{\BIBentrySTDinterwordspacing}{\spaceskip=0pt\relax}
\providecommand{\BIBentryALTinterwordstretchfactor}{4}
\providecommand{\BIBentryALTinterwordspacing}{\spaceskip=\fontdimen2\font plus
\BIBentryALTinterwordstretchfactor\fontdimen3\font minus \fontdimen4\font\relax}
\providecommand{\BIBforeignlanguage}[2]{{%
\expandafter\ifx\csname l@#1\endcsname\relax
\typeout{** WARNING: IEEEtran.bst: No hyphenation pattern has been}%
\typeout{** loaded for the language `#1'. Using the pattern for}%
\typeout{** the default language instead.}%
\else
\language=\csname l@#1\endcsname
\fi
#2}}
\providecommand{\BIBdecl}{\relax}
\BIBdecl

\bibitem{preskill2018quantum}
J.~Preskill, ``Quantum computing in the nisq era and beyond,'' \emph{Quantum}, vol.~2, p.~79, 2018.

\bibitem{bauer2020quantum}
B.~Bauer, S.~Bravyi, M.~Motta, and G.~K.-L. Chan, ``Quantum algorithms for quantum chemistry and quantum materials science,'' \emph{Chemical Reviews}, vol. 120, no.~22, pp. 12\,685--12\,717, 2020.

\bibitem{liang2022variational}
Z.~Liang, H.~Wang, J.~Cheng, Y.~Ding, H.~Ren, Z.~Gao, Z.~Hu, D.~S. Boning, X.~Qian, S.~Han \emph{et~al.}, ``Variational quantum pulse learning,'' in \emph{2022 IEEE International Conference on Quantum Computing and Engineering (QCE)}.\hskip 1em plus 0.5em minus 0.4em\relax IEEE, 2022, pp. 556--565.

\bibitem{thanasilp2023subtleties}
S.~Thanasilp, S.~Wang, N.~A. Nghiem, P.~Coles, and M.~Cerezo, ``Subtleties in the trainability of quantum machine learning models,'' \emph{Quantum Machine Intelligence}, vol.~5, no.~1, p.~21, 2023.

\bibitem{zhang2024generative}
B.~Zhang, P.~Xu, X.~Chen, and Q.~Zhuang, ``Generative quantum machine learning via denoising diffusion probabilistic models,'' \emph{Physical Review Letters}, vol. 132, no.~10, p. 100602, 2024.

\bibitem{vijendran2024expressive}
V.~Vijendran, A.~Das, D.~E. Koh, S.~M. Assad, and P.~K. Lam, ``An expressive ansatz for low-depth quantum approximate optimisation,'' \emph{Quantum Science and Technology}, vol.~9, no.~2, p. 025010, 2024.

\bibitem{chiribella2008quantum}
G.~Chiribella, G.~M. D’Ariano, and P.~Perinotti, ``Quantum circuit architecture,'' \emph{Physical review letters}, vol. 101, no.~6, p. 060401, 2008.

\bibitem{liang2023unleashing}
Z.~Liang, J.~Cheng, R.~Yang, H.~Ren, Z.~Song, D.~Wu, X.~Qian, T.~Li, and Y.~Shi, ``Unleashing the potential of llms for quantum computing: A study in quantum architecture design,'' \emph{arXiv preprint arXiv:2307.08191}, 2023.

\bibitem{cerezo2021cost}
M.~Cerezo, A.~Sone, T.~Volkoff, L.~Cincio, and P.~J. Coles, ``Cost function dependent barren plateaus in shallow parametrized quantum circuits,'' \emph{Nature communications}, vol.~12, no.~1, p. 1791, 2021.

\bibitem{uvarov2021barren}
A.~Uvarov and J.~D. Biamonte, ``On barren plateaus and cost function locality in variational quantum algorithms,'' \emph{Journal of Physics A: Mathematical and Theoretical}, vol.~54, no.~24, p. 245301, 2021.

\bibitem{McClean2018landscapes}
J.~R. McClean, S.~Boixo, V.~N. Smelyanskiy, R.~Babbush, and H.~Neven, ``Barren plateaus in quantum neural network training landscapes,'' \emph{Nature communications}, vol.~9, no.~1, p. 4812, 2018.

\bibitem{qi2023barren}
H.~Qi, L.~Wang, H.~Zhu, A.~Gani, and C.~Gong, ``The barren plateaus of quantum neural networks: review, taxonomy and trends,'' \emph{Quantum Information Processing}, vol.~22, no.~12, p. 435, 2023.

\bibitem{renes2004symmetric}
J.~M. Renes, R.~Blume-Kohout, A.~J. Scott, and C.~M. Caves, ``Symmetric informationally complete quantum measurements,'' \emph{Journal of Mathematical Physics}, vol.~45, no.~6, pp. 2171--2180, 2004.

\bibitem{wang2021noise}
S.~Wang, E.~Fontana, K.~Sharma, A.~Sone, L.~Cincio, and P.~J. Coles, ``Noise-induced barren plateaus in variational quantum algorithms,'' \emph{Nature communications}, vol.~12, no.~1, p. 6961, 2021.

\bibitem{singkanipa2024beyond}
P.~Singkanipa and D.~A. Lidar, ``Beyond unital noise in variational quantum algorithms: noise-induced barren plateaus and fixed points,'' \emph{arXiv preprint arXiv:2402.08721}, 2024.

\bibitem{marrero2021entanglement}
C.~O. Marrero, M.~Kieferov{\'a}, and N.~Wiebe, ``Entanglement-induced barren plateaus,'' \emph{PRX Quantum}, vol.~2, no.~4, p. 040316, 2021.

\bibitem{holmes2021barren}
Z.~Holmes, A.~Arrasmith, B.~Yan, P.~J. Coles, A.~Albrecht, and A.~T. Sornborger, ``Barren plateaus preclude learning scramblers,'' \emph{Physical Review Letters}, vol. 126, no.~19, p. 190501, 2021.

\bibitem{holmes2022connecting}
Z.~Holmes, K.~Sharma, M.~Cerezo, and P.~J. Coles, ``Connecting ansatz expressibility to gradient magnitudes and barren plateaus,'' \emph{PRX Quantum}, vol.~3, no.~1, p. 010313, 2022.

\bibitem{ragone2023unified}
M.~Ragone, B.~N. Bakalov, F.~Sauvage, A.~F. Kemper, C.~O. Marrero, M.~Larocca, and M.~Cerezo, ``A unified theory of barren plateaus for deep parametrized quantum circuits,'' \emph{arXiv preprint arXiv:2309.09342}, 2023.

\bibitem{liu2024laziness}
J.~Liu, Z.~Lin, and L.~Jiang, ``Laziness, barren plateau, and noises in machine learning,'' \emph{Machine Learning: Science and Technology}, vol.~5, no.~1, p. 015058, 2024.

\bibitem{friedrich2024barren}
L.~Friedrich, T.~d.~S. Farias, and J.~Maziero, ``Barren plateaus induced by the dimension of qudits,'' \emph{arXiv preprint arXiv:2405.08190}, 2024.

\bibitem{fontana2024adjoint}
E.~Fontana, D.~Herman, S.~Chakrabarti, N.~Kumar, R.~Yalovetzky, J.~Heredge, S.~H. Sureshbabu, and M.~Pistoia, ``The adjoint is all you need: Characterizing barren plateaus in quantum ans{\"a}tze,'' \emph{Bulletin of the American Physical Society}, 2024.

\bibitem{diaz2023showcasing}
N.~Diaz, D.~Garc{\'\i}a-Mart{\'\i}n, S.~Kazi, M.~Larocca, and M.~Cerezo, ``Showcasing a barren plateau theory beyond the dynamical lie algebra,'' \emph{arXiv preprint arXiv:2310.11505}, 2023.

\bibitem{goh2023lie}
\BIBentryALTinterwordspacing
M.~L. Goh, M.~Larocca, L.~Cincio, M.~Cerezo, and F.~Sauvage, ``Lie-algebraic classical simulations for variational quantum computing,'' \emph{arXiv preprint arXiv:2308.01432}, 2023. [Online]. Available: \url{https://arxiv.org/abs/2308.01432}
\BIBentrySTDinterwordspacing

\bibitem{patti2021entanglement}
T.~L. Patti, K.~Najafi, X.~Gao, and S.~F. Yelin, ``Entanglement devised barren plateau mitigation,'' \emph{Physical Review Research}, vol.~3, no.~3, p. 033090, 2021.

\bibitem{anschuetz2022quantum}
E.~R. Anschuetz and B.~T. Kiani, ``Quantum variational algorithms are swamped with traps,'' \emph{Nature Communications}, vol.~13, no.~1, p. 7760, 2022.

\bibitem{wiersema2020exploring}
R.~Wiersema, C.~Zhou, Y.~de~Sereville, J.~F. Carrasquilla, Y.~B. Kim, and H.~Yuen, ``Exploring entanglement and optimization within the hamiltonian variational ansatz,'' \emph{PRX Quantum}, vol.~1, no.~2, p. 020319, 2020.

\bibitem{mao2023barren}
R.~Mao, G.~Tian, and X.~Sun, ``Barren plateaus of alternated disentangled ucc ansatzs,'' \emph{arXiv preprint arXiv:2312.08105}, 2023.

\bibitem{liu2022presence}
Z.~Liu, L.-W. Yu, L.-M. Duan, and D.-L. Deng, ``Presence and absence of barren plateaus in tensor-network based machine learning,'' \emph{Physical Review Letters}, vol. 129, no.~27, p. 270501, 2022.

\bibitem{martin2023barren}
E.~C. Mart{\'\i}n, K.~Plekhanov, and M.~Lubasch, ``Barren plateaus in quantum tensor network optimization,'' \emph{Quantum}, vol.~7, p. 974, 2023.

\bibitem{cybulski2023impact}
J.~L. Cybulski and T.~Nguyen, ``Impact of barren plateaus countermeasures on the quantum neural network capacity to learn,'' \emph{Quantum Information Processing}, vol.~22, no.~12, p. 442, 2023.

\bibitem{abbas2021power}
A.~Abbas, D.~Sutter, C.~Zoufal, A.~Lucchi, A.~Figalli, and S.~Woerner, ``The power of quantum neural networks,'' \emph{Nature Computational Science}, vol.~1, no.~6, pp. 403--409, 2021.

\bibitem{pesah2021absence}
A.~Pesah, M.~Cerezo, S.~Wang, T.~Volkoff, A.~T. Sornborger, and P.~J. Coles, ``Absence of barren plateaus in quantum convolutional neural networks,'' \emph{Physical Review X}, vol.~11, no.~4, p. 041011, 2021.

\bibitem{coelho2024vqc}
R.~Coelho, A.~Sequeira, and L.~P. Santos, ``Vqc-based reinforcement learning with data re-uploading: Performance and trainability,'' \emph{arXiv preprint arXiv:2401.11555}, 2024.

\bibitem{zhao2021analyzing}
C.~Zhao and X.-S. Gao, ``Analyzing the barren plateau phenomenon in training quantum neural networks with the zx-calculus,'' \emph{Quantum}, vol.~5, p. 466, 2021.

\bibitem{cerezo2021higher}
M.~Cerezo and P.~J. Coles, ``Higher order derivatives of quantum neural networks with barren plateaus,'' \emph{Quantum Science and Technology}, vol.~6, no.~3, p. 035006, 2021.

\bibitem{cerezo2023does}
M.~Cerezo, M.~Larocca, D.~Garc{\'\i}a-Mart{\'\i}n, N.~Diaz, P.~Braccia, E.~Fontana, M.~S. Rudolph, P.~Bermejo, A.~Ijaz, S.~Thanasilp \emph{et~al.}, ``Does provable absence of barren plateaus imply classical simulability? or, why we need to rethink variational quantum computing,'' \emph{arXiv preprint arXiv:2312.09121}, 2023.

\bibitem{arrasmith2021effect}
A.~Arrasmith, M.~Cerezo, P.~Czarnik, L.~Cincio, and P.~J. Coles, ``Effect of barren plateaus on gradient-free optimization,'' \emph{Quantum}, vol.~5, p. 558, 2021.

\bibitem{arrasmith2022equivalence}
A.~Arrasmith, Z.~Holmes, M.~Cerezo, and P.~J. Coles, ``Equivalence of quantum barren plateaus to cost concentration and narrow gorges,'' \emph{Quantum Science and Technology}, vol.~7, no.~4, p. 045015, 2022.

\bibitem{miao2024equivalence}
Q.~Miao and T.~Barthel, ``Equivalence of cost concentration and gradient vanishing for quantum circuits: an elementary proof in the riemannian formulation,'' \emph{Quantum Science and Technology}, vol.~9, no.~4, 2024.

\bibitem{barthel2023absence}
T.~Barthel and Q.~Miao, ``Absence of barren plateaus and scaling of gradients in the energy optimization of isometric tensor network states,'' \emph{arXiv preprint arXiv:2304.00161}, 2023.

\bibitem{miao2024isometric}
Q.~Miao and T.~Barthel, ``Isometric tensor network optimization for extensive hamiltonians is free of barren plateaus,'' \emph{Physical Review A}, vol. 109, no.~5, p. L050402, 2024.

\bibitem{cao2024exploiting}
C.~Cao, Y.~Zhou, S.~Tannu, N.~Shannon, and R.~Joynt, ``Exploiting many-body localization for scalable variational quantum simulation,'' \emph{arXiv preprint arXiv:2404.17560}, 2024.

\bibitem{perez2024analyzing}
A.~P{\'e}rez-Salinas, H.~Wang, and X.~Bonet-Monroig, ``Analyzing variational quantum landscapes with information content,'' \emph{npj Quantum Information}, vol.~10, no.~1, p.~27, 2024.

\bibitem{nemkov2024barren}
N.~A. Nemkov, E.~O. Kiktenko, and A.~K. Fedorov, ``Barren plateaus are swamped with traps,'' \emph{arXiv preprint arXiv:2405.05332}, 2024.

\bibitem{letcher2023tight}
A.~Letcher, S.~Woerner, and C.~Zoufal, ``From tight gradient bounds for parameterized quantum circuits to the absence of barren plateaus in qgans,'' \emph{arXiv preprint arXiv:2309.12681}, 2023.

\bibitem{larocca2022diagnosing}
M.~Larocca, P.~Czarnik, K.~Sharma, G.~Muraleedharan, P.~J. Coles, and M.~Cerezo, ``Diagnosing barren plateaus with tools from quantum optimal control,'' \emph{Quantum}, vol.~6, p. 824, 2022.

\bibitem{park2024quantum}
S.~Park and J.~Kim, ``Quantum neural network software testing, analysis, and code optimization for advanced iot systems: Design, implementation, and visualization,'' \emph{arXiv preprint arXiv:2401.10914}, 2024.

\bibitem{kashif2024alleviating}
M.~Kashif, M.~Rashid, S.~Al-Kuwari, and M.~Shafique, ``Alleviating barren plateaus in parameterized quantum machine learning circuits: Investigating advanced parameter initialization strategies,'' in \emph{2024 Design, Automation \& Test in Europe Conference \& Exhibition (DATE)}.\hskip 1em plus 0.5em minus 0.4em\relax IEEE, 2024, pp. 1--6.

\bibitem{grant2019initialization}
E.~Grant, L.~Wossnig, M.~Ostaszewski, and M.~Benedetti, ``An initialization strategy for addressing barren plateaus in parametrized quantum circuits,'' \emph{Quantum}, vol.~3, p. 214, 2019.

\bibitem{sauvage2021flip}
F.~Sauvage, S.~Sim, A.~A. Kunitsa, W.~A. Simon, M.~Mauri, and A.~Perdomo-Ortiz, ``Flip: A flexible initializer for arbitrarily-sized parametrized quantum circuits,'' \emph{arXiv preprint arXiv:2103.08572}, 2021.

\bibitem{sack2022avoiding}
S.~H. Sack, R.~A. Medina, A.~A. Michailidis, R.~Kueng, and M.~Serbyn, ``Avoiding barren plateaus using classical shadows,'' \emph{PRX Quantum}, vol.~3, no.~2, p. 020365, 2022.

\bibitem{rad2022surviving}
A.~Rad, A.~Seif, and N.~M. Linke, ``Surviving the barren plateau in variational quantum circuits with bayesian learning initialization,'' \emph{arXiv preprint arXiv:2203.02464}, 2022.

\bibitem{kulshrestha2022beinit}
A.~Kulshrestha and I.~Safro, ``Beinit: Avoiding barren plateaus in variational quantum algorithms,'' in \emph{2022 IEEE international conference on quantum computing and engineering (QCE)}.\hskip 1em plus 0.5em minus 0.4em\relax IEEE, 2022, pp. 197--203.

\bibitem{zhang2022escaping}
K.~Zhang, L.~Liu, M.-H. Hsieh, and D.~Tao, ``Escaping from the barren plateau via gaussian initializations in deep variational quantum circuits,'' \emph{Advances in Neural Information Processing Systems}, vol.~35, pp. 18\,612--18\,627, 2022.

\bibitem{friedrich2022avoiding}
L.~Friedrich and J.~Maziero, ``Avoiding barren plateaus with classical deep neural networks,'' \emph{Physical Review A}, vol. 106, no.~4, p. 042433, 2022.

\bibitem{mele2022avoiding}
A.~A. Mele, G.~B. Mbeng, G.~E. Santoro, M.~Collura, and P.~Torta, ``Avoiding barren plateaus via transferability of smooth solutions in a hamiltonian variational ansatz,'' \emph{Physical Review A}, vol. 106, no.~6, p. L060401, 2022.

\bibitem{grimsley2023adaptive}
H.~R. Grimsley, G.~S. Barron, E.~Barnes, S.~E. Economou, and N.~J. Mayhall, ``Adaptive, problem-tailored variational quantum eigensolver mitigates rough parameter landscapes and barren plateaus,'' \emph{npj Quantum Information}, vol.~9, no.~1, p.~19, 2023.

\bibitem{liu2023mitigating}
H.-Y. Liu, T.-P. Sun, Y.-C. Wu, Y.-J. Han, and G.-P. Guo, ``Mitigating barren plateaus with transfer-learning-inspired parameter initializations,'' \emph{New Journal of Physics}, vol.~25, no.~1, p. 013039, 2023.

\bibitem{park2024hamiltonian}
C.-Y. Park and N.~Killoran, ``Hamiltonian variational ansatz without barren plateaus,'' \emph{Quantum}, vol.~8, p. 1239, 2024.

\bibitem{kempe2006complexity}
J.~Kempe, A.~Kitaev, and O.~Regev, ``The complexity of the local hamiltonian problem,'' \emph{Siam journal on computing}, vol.~35, no.~5, pp. 1070--1097, 2006.

\bibitem{cubitt2016complexity}
T.~Cubitt and A.~Montanaro, ``Complexity classification of local hamiltonian problems,'' \emph{SIAM Journal on Computing}, vol.~45, no.~2, pp. 268--316, 2016.

\bibitem{anshu2016concentration}
A.~Anshu, ``Concentration bounds for quantum states with finite correlation length on quantum spin lattice systems,'' \emph{New Journal of Physics}, vol.~18, no.~8, p. 083011, 2016.

\bibitem{park2024hardware}
C.-Y. Park, M.~Kang, and J.~Huh, ``Hardware-efficient ansatz without barren plateaus in any depth,'' \emph{arXiv preprint arXiv:2403.04844}, 2024.

\bibitem{nadori2024promising}
J.~N{\'a}dori, G.~Morse, Z.~Majnay-Tak{\'a}cs, Z.~Zimbor{\'a}s, and P.~Rakyta, ``The promising path of evolutionary optimization to avoid barren plateaus,'' \emph{arXiv preprint arXiv:2402.05227}, 2024.

\bibitem{ostaszewski2021structure}
M.~Ostaszewski, E.~Grant, and M.~Benedetti, ``Structure optimization for parameterized quantum circuits,'' \emph{Quantum}, vol.~5, p. 391, 2021.

\bibitem{skolik2021layerwise}
A.~Skolik, J.~R. McClean, M.~Mohseni, P.~Van Der~Smagt, and M.~Leib, ``Layerwise learning for quantum neural networks,'' \emph{Quantum Machine Intelligence}, vol.~3, pp. 1--11, 2021.

\bibitem{gharibyan2023hierarchical}
H.~Gharibyan, V.~Su, and H.~Tepanyan, ``Hierarchical learning for quantum ml: Novel training technique for large-scale variational quantum circuits,'' \emph{arXiv preprint arXiv:2311.12929}, 2023.

\bibitem{liu2024mitigating}
X.~Liu, G.~Liu, H.-K. Zhang, J.~Huang, and X.~Wang, ``Mitigating barren plateaus of variational quantum eigensolvers,'' \emph{IEEE Transactions on Quantum Engineering}, 2024.

\bibitem{haug2021optimal}
T.~Haug and M.~Kim, ``Optimal training of variational quantum algorithms without barren plateaus,'' \emph{arXiv preprint arXiv:2104.14543}, 2021.

\bibitem{kieferova2021quantum}
M.~Kieferova, O.~M. Carlos, and N.~Wiebe, ``Quantum generative training using r$\backslash$'enyi divergences,'' \emph{arXiv preprint arXiv:2106.09567}, 2021.

\bibitem{heyraud2023efficient}
V.~Heyraud, Z.~Li, K.~Donatella, A.~Le~Boit{\'e}, and C.~Ciuti, ``Efficient estimation of trainability for variational quantum circuits,'' \emph{PRX Quantum}, vol.~4, no.~4, p. 040335, 2023.

\bibitem{sannia2023engineered}
A.~Sannia, F.~Tacchino, I.~Tavernelli, G.~L. Giorgi, and R.~Zambrini, ``Engineered dissipation to mitigate barren plateaus,'' \emph{arXiv preprint arXiv:2310.15037}, 2023.

\bibitem{mele2024noise}
A.~A. Mele, A.~Angrisani, S.~Ghosh, S.~Khatri, J.~Eisert, D.~S. Fran{\c{c}}a, and Y.~Quek, ``Noise-induced shallow circuits and absence of barren plateaus,'' \emph{arXiv preprint arXiv:2403.13927}, 2024.

\bibitem{broers2024mitigated}
L.~Broers and L.~Mathey, ``Mitigated barren plateaus in the time-nonlocal optimization of analog quantum-algorithm protocols,'' \emph{Physical Review Research}, vol.~6, no.~1, p. 013076, 2024.

\bibitem{zambrano2024avoiding}
L.~Zambrano, A.~D. Mu{\~n}oz-Moller, M.~Mu{\~n}oz, L.~Pereira, and A.~Delgado, ``Avoiding barren plateaus in the variational determination of geometric entanglement,'' \emph{Quantum Science and Technology}, vol.~9, no.~2, p. 025016, 2024.

\bibitem{sciorilli2024towards}
M.~Sciorilli, L.~Borges, T.~L. Patti, D.~Garc{\'\i}a-Mart{\'\i}n, G.~Camilo, A.~Anandkumar, and L.~Aolita, ``Towards large-scale quantum optimization solvers with few qubits,'' \emph{arXiv preprint arXiv:2401.09421}, 2024.

\bibitem{falla2024graph}
J.~Falla, Q.~Langfitt, Y.~Alexeev, and I.~Safro, ``Graph representation learning for parameter transferability in quantum approximate optimization algorithm,'' \emph{arXiv preprint arXiv:2401.06655}, 2024.

\bibitem{yao2024avoiding}
Y.~Yao and Y.~Hasegawa, ``Avoiding barren plateaus with entanglement,'' \emph{arXiv preprint arXiv:2406.03748}, 2024.

\bibitem{friedrich2024quantum}
L.~Friedrich and J.~Maziero, ``Quantum neural network with ensemble learning to mitigate barren plateaus and cost function concentration,'' \emph{arXiv preprint arXiv:2402.06026}, 2024.

\bibitem{shi2024avoiding}
X.~Shi and Y.~Shang, ``Avoiding barren plateaus via gaussian mixture model,'' \emph{arXiv preprint arXiv:2402.13501}, 2024.

\bibitem{li2021vsql}
G.~Li, Z.~Song, and X.~Wang, ``Vsql: Variational shadow quantum learning for classification,'' in \emph{Proceedings of the AAAI conference on artificial intelligence}, vol.~35, no.~9, 2021, pp. 8357--8365.

\bibitem{bharti2021quantum}
K.~Bharti and T.~Haug, ``Quantum-assisted simulator,'' \emph{Physical Review A}, vol. 104, no.~4, p. 042418, 2021.

\bibitem{du2022quantum}
Y.~Du, T.~Huang, S.~You, M.-H. Hsieh, and D.~Tao, ``Quantum circuit architecture search for variational quantum algorithms,'' \emph{npj Quantum Information}, vol.~8, no.~1, p.~62, 2022.

\bibitem{zhang2022quark}
Z.~Zhang, Z.~Chen, H.~Huang, and Z.~Jia, ``Quark: A gradient-free quantum learning framework for classification tasks,'' \emph{arXiv preprint arXiv:2210.01311}, 2022.

\bibitem{selvarajan2023dimensionality}
R.~Selvarajan, M.~Sajjan, T.~S. Humble, and S.~Kais, ``Dimensionality reduction with variational encoders based on subsystem purification,'' \emph{Mathematics}, vol.~11, no.~22, p. 4678, 2023.

\bibitem{tuysuz2023classical}
C.~T{\"u}ys{\"u}z, G.~Clemente, A.~Crippa, T.~Hartung, S.~K{\"u}hn, and K.~Jansen, ``Classical splitting of parametrized quantum circuits,'' \emph{Quantum Machine Intelligence}, vol.~5, no.~2, p.~34, 2023.

\bibitem{kashif2024resqnets}
M.~Kashif and S.~Al-Kuwari, ``Resqnets: a residual approach for mitigating barren plateaus in quantum neural networks,'' \emph{EPJ Quantum Technology}, vol.~11, no.~1, p.~4, 2024.

\bibitem{shin2024layerwise}
M.~Shin, S.~Lee, M.~Lee, D.~Ji, H.~Yeo, H.~J. Lee, and K.~Jeong, ``Layerwise quantum convolutional neural networks provide a unified way for estimating fundamental properties of quantum information theory,'' \emph{arXiv preprint arXiv:2401.07716}, 2024.

\bibitem{zhang2024absence}
H.-K. Zhang, S.~Liu, and S.-X. Zhang, ``Absence of barren plateaus in finite local-depth circuits with long-range entanglement,'' \emph{Physical Review Letters}, vol. 132, no.~15, p. 150603, 2024.

\bibitem{zhuang2024improving}
J.~Zhuang, J.~Cunningham, and C.~Guan, ``Improving trainability of variational quantum circuits via regularization strategies,'' \emph{arXiv preprint arXiv:2405.01606}, 2024.

\bibitem{rappaport2023measurement}
S.~Rappaport, G.~Gyawali, T.~Sereno, and M.~J. Lawler, ``Measurement-induced landscape transitions in hybrid variational quantum circuits,'' \emph{arXiv preprint arXiv:2312.09135}, 2023.

\bibitem{larocca2024review}
M.~Larocca, S.~Thanasilp, S.~Wang, K.~Sharma, J.~Biamonte, P.~J. Coles, L.~Cincio, J.~R. McClean, Z.~Holmes, and M.~Cerezo, ``A review of barren plateaus in variational quantum computing,'' \emph{arXiv preprint arXiv:2405.00781}, 2024.

\bibitem{gelman2024survey}
M.~Gelman, ``A survey of methods for mitigating barren plateaus for parameterized quantum circuits,'' \emph{arXiv preprint arXiv:2406.14285}, 2024.

\end{thebibliography}
\balance
\clearpage

\end{document}